   \documentclass{aa}  

\usepackage{graphicx}
\usepackage{txfonts}
\newcommand{\MJD}[1]{MJD~#1}
\newcommand{\mum}{\mu{\rm m}}

\begin{document} 

\title{Evolving optical polarisation of the black hole X-ray binary MAXI~J1820+070}
\titlerunning{Polarisation of MAXI~J1820+070}

\author{Alexandra Veledina\inst{1,2,3} \and Andrei V. Berdyugin\inst{1} \and Ilia A. Kosenkov\inst{1,4} \and Jari J. E. Kajava\inst{5} \and Sergey S. Tsygankov\inst{1,3} \and Vilppu Piirola\inst{1} \and Svetlana V. Berdyugina\inst{6,7} \and Takeshi Sakanoi\inst{8} \and Masato Kagitani\inst{8} \and Vadim Kravtsov\inst{4} \and Juri Poutanen\inst{1,2,3}
          }
\authorrunning{Veledina et al.} 

\institute{
Department of Physics and Astronomy, FI-20014 University of Turku, Finland 
\\ \email{alexandra.veledina@gmail.com} 
\and Nordita, KTH Royal Institute of Technology and Stockholm University, Roslagstullsbacken 23, SE-10691 Stockholm, Sweden
\and Space Research Institute of the Russian Academy of Sciences, Profsoyuznaya Str. 84/32, 117997 Moscow,  Russia
\and Department of Astrophysics, St. Petersburg State University, Universitetskiy pr. 28, Peterhof, 198504 St. Petersburg, Russia
\and Finnish Centre for Astronomy with ESO (FINCA), FI-20014 University of Turku, Finland
\and Kiepenheuer-Institute f\"{u}r Sonnenphysik, Sch\"{o}neckstr. 6, 79104 Freiburg, Germany
\and Institute for Astronomy, University of Hawaii, 2680 Woodlawn Drive, Honolulu, 96822-1897 HI, USA
\and Graduate School of Science, Tohoku University, Aoba-ku, 980-8578 Sendai, Japan
             }

   \date{Received September 15, 1996; accepted March 16, 1997}

  \abstract
   {}
{The optical emission of black hole transients increases by several magnitudes during the X-ray outbursts. 
Whether the extra light arises from the X-ray heated outer disc, from the inner hot accretion flow, or from the jet is currently debated. 
Optical polarisation measurements are able to distinguish the relative contributions of these components.}
   {We present the results of {\it BVR} polarisation
   measurements of the black hole X-ray binary MAXI~J1820+070 during the period of March-April 2018.}
   {We detect small, $\sim$0.7\%, 
   but statistically significant polarisation, part of which is of interstellar origin.
Depending on the interstellar polarisation estimate, the intrinsic polarisation degree of the source is between $\sim$0.3\% and 0.7\%, and the polarisation position angle is between $\sim$$10\degr-30$$\degr$.
We show that the polarisation increases after MJD~58222 (2018 April 14). 
The change is of the order of 0.1\% and is most pronounced in the {\it R} band.
The change of the source Stokes parameters occurs simultaneously with the drop of the observed {\it V}-band flux and a slow softening of the X-ray spectrum.
The Stokes vectors of intrinsic polarisation before and after the drop are parallel, at least in the {\it V} and {\it R} filters. }
   {We suggest that the increased polarisation is due to the decreasing contribution of the non-polarized component, which we associate with the the hot flow or jet emission. 
The low polarisation can result from the tangled geometry of the magnetic field or from the Faraday rotation in the dense, ionised, and magnetised medium close to the black hole. 
The polarized optical emission is likely produced by the irradiated disc or  by scattering of its radiation in the optically thin outflow.}


   \maketitle
%
%________________________________________________________________

\section{Introduction}

The black hole X-ray binaries play a key role in understanding the processes of accretion onto and ejection from the compact objects in the presence of a strong gravitational field.
These sources evolve through the cycle of the bright outbursts that proceed on timescales of weeks to months, separated by long periods of quiescence of years to decades \citep[see][for reviews]{RM06,DGK07}. 
Their typical X-ray luminosities at the outburst peak are $\sim$$10^{37}-10^{39}$~erg~s$^{-1}$, and typical distances are a few kpc.
A few exceptional sources have reached or exceeded fluxes of 1~Crab in the X-rays: A~0620--00 \citep{Kuulkers98},
V404~Cyg \citep{Makino1989,Rodriguez15,Motta17} and V4641~Sgr \citep{Hjellming00,Revnivtsev02}.

The X-ray transient MAXI~J1820+070 was first detected on 2018 March 11 \citep{ATel11399} by the Monitor of All-sky X-ray Image \citep[MAXI,][]{MAXI09} and was associated with the optical transient ASASSN-18ey \citep{ATel11400,Tucker18}.
In the X-rays, the source flux exceeded 3~Crabs \citep{ATel11478,ATel11488}, and in the optical, the source reached a magnitude of $m_{V}=12-13$ \citep{ATel11421,ATel11533}.  
The parallax of the source $\pi=0.3\pm0.1$~mas was presented in the {\it Gaia} DR2 catalogue \citep{GaiaDR2}. 
This corresponds to a distance of $3.9^{+3.3}_{-1.3}$~kpc \citep{GRJ18}.  
This unusually bright event allows a detailed investigation of multi-wavelength spectral and timing properties. 
The course of the outburst was monitored in radio \citep{ATel11539,ATel11540}, sub-millimeter \citep{ATel11440}, optical \citep{ATel11418}, X-rays \citep{ATel11423}, and $\gamma$-rays \citep{ATel11478,ATel11490}.
Because the object was sufficiently bright even for small telescopes, the target was almost continuously monitored, and a rich variety of phenomena was observed.
Fast variability and powerful flares in the optical and infrared \citep{ATel11421,ATel11426,ATel11437,ATel11451}, optical, and X-ray quasi-periodic oscillations \citep{ATel11488,ATel11510,ATel11578,ATel11591,ATel11723} as well as low linear polarisation \citep{ATel11445} were detected in the source.
A 17~h photometric period was recently reported \citep{ATel11756} and was tentatively associated with the orbital or superhump period (previously, the source showed a 3.4~h periodicity, \citealt{ATel11596}).
The X-ray spectral and timing properties as well as the optical-to-X-ray flux ratio suggests that the source is a black hole binary \citep{ATel11418,ATel11488}.

The origin of the optical emission of black hole transients in the outburst is still debated \citep[see][]{PV14}.  
The optical flux can be produced in the outer disc that is irradiated by the central X-ray source, in the jet, and in the inner hot accretion flow. 
Accurate measurements of optical polarisation at different outburst stages with simultaneous studies of optical and X-ray spectral and timing properties are expected to  help separate the relative contributions of these components.
In this work, we present the results of our polarisation campaign with the highly sensitive Dipol-2 instrument \citep{PBB14}, which was conducted at the initial stages of the outburst. 
We show that the source demonstrates small but statistically significant intrinsic polarisation, and we study its spectral and temporal properties.
We describe the data in Sect.~\ref{sect:data}, present the results in Sect.~\ref{sect:results}, discuss them within the  accretion-ejection framework in Sect.~\ref{sect:discuss}, and summarise our findings in Sect.~\ref{sect:conclus}.

% fig 1 
 \begin{figure}
\center{\includegraphics[width=0.8\columnwidth]{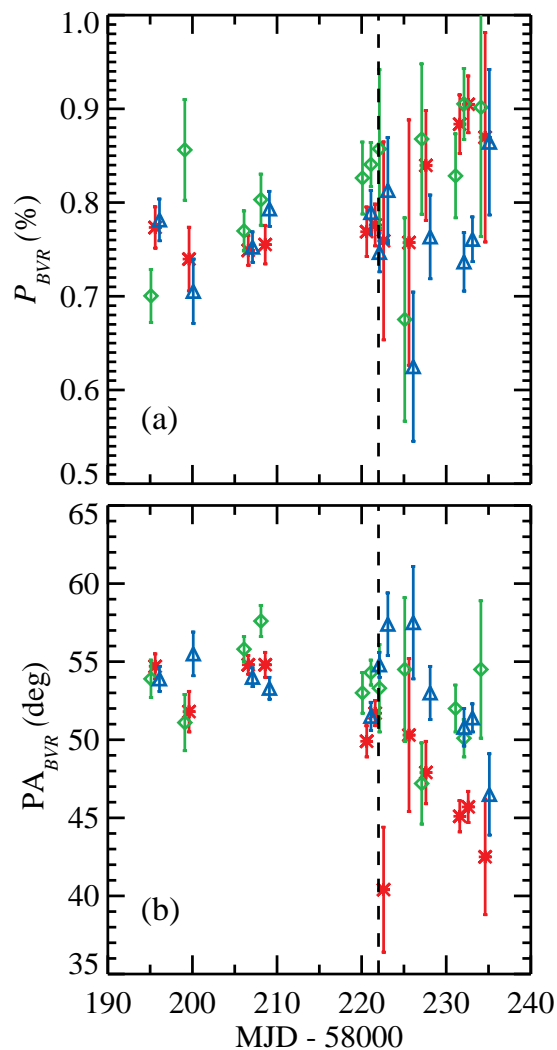}}
     \caption{Evolution of the polarisation degree and polarisation angle of MAXI~J1820+070 in different filters: $B$ (blue triangles), $V$ (green diamonds), and $R$ (red crosses). For clarity, 0.5~d was added to MJD of the $B$ filter and subtracted from MJD of the $V$ filter. The vertical dashed line marks MJD~58222.  
     }  
    \label{fig:lc_scale}
 \end{figure}

% Table 1 source 
\begin{table*}
        \centering 
        \caption{Polarimetric data for MAXI~J1820+070. }
        \begin{tabular}{ccccccc}
                \hline
        &       \multicolumn{2}{c}{$B$} & \multicolumn{2}{c}{$V$} & \multicolumn{2}{c}{$R$} \\
  MJD   &       $P$ (\%) & PA ($\degr$) & $P$ (\%) & PA ($\degr$) & $P$ (\%) & PA ($\degr$)\\ 
                \hline     
        58195.61983   &   $0.78  \pm  0.02$   &    $53.9  \pm  0.8$   &   $0.70  \pm  0.03$   &    $53.9  \pm  1.2$   &   $0.77  \pm  0.02$   &   $54.7  \pm  0.8$   \\
        58199.61832   &   $0.71  \pm  0.03$   &    $55.5  \pm  1.4$   &   $0.86  \pm  0.05$   &    $51.1  \pm  1.8$   &   $0.74  \pm  0.03$   &   $51.8  \pm  1.3$   \\
        58206.61784   &   $0.75  \pm  0.02$   &    $54.0  \pm  0.6$   &   $0.77  \pm  0.02$   &    $55.8  \pm  0.8$   &   $0.75  \pm  0.02$   &   $54.8  \pm  0.6$   \\
        58208.61502   &   $0.79  \pm  0.02$   &    $53.3  \pm  0.7$   &   $0.80  \pm  0.03$   &    $57.6  \pm  1.0$   &   $0.76  \pm  0.02$   &   $54.8  \pm  0.8$   \\
        58220.61279   &   $0.79  \pm  0.02$   &    $51.5  \pm  0.9$   &   $0.83  \pm  0.04$   &    $53.0  \pm  1.3$   &   $0.77  \pm  0.03$   &   $49.9  \pm  1.0$   \\
        58221.59912   &   $0.75  \pm  0.02$   &    $54.8  \pm  0.8$   &   $0.84  \pm  0.02$   &    $54.3  \pm  0.8$   &   $0.78  \pm  0.02$   &   $51.7  \pm  0.8$   \\
        58222.58917   &   $0.81  \pm  0.06$   &    $57.4  \pm  2.0$   &   $0.86  \pm  0.08$   &    $53.3  \pm  2.8$   &   $0.76  \pm  0.11$   &   $40.4  \pm  4.0$   \\
        58225.59632   &   $0.62  \pm  0.08$   &    $57.5  \pm  3.6$   &   $0.68  \pm  0.11$   &    $54.5  \pm  4.6$   &   $0.76  \pm  0.13$   &   $50.3  \pm  4.9$   \\
        58227.57404   &   $0.76  \pm  0.04$   &    $53.0  \pm  1.7$   &   $0.87  \pm  0.08$   &    $47.2  \pm  2.6$   &   $0.84  \pm  0.06$   &   $47.9  \pm  2.0$   \\
        58231.60084   &   $0.74  \pm  0.03$   &    $50.8  \pm  1.2$   &   $0.83  \pm  0.04$   &    $52.0  \pm  1.5$   &   $0.88  \pm  0.03$   &   $45.1  \pm  1.0$   \\
        58232.60051   &   $0.76  \pm  0.02$   &    $51.4  \pm  0.9$   &   $0.91  \pm  0.04$   &    $50.1  \pm  1.2$   &   $0.90  \pm  0.03$  &   $45.7  \pm  1.0$   \\
        58234.60428   &   $0.86  \pm  0.08$   &    $46.5  \pm  2.6$   &   $0.9  \pm  0.14$   &    $54.5  \pm  4.4$   &   $0.87  \pm  0.10$   &   $42.5  \pm  3.7$   \\
\hline
        $<$58222   &   $0.76  \pm  0.01$   &    $53.9  \pm  0.3$   &   $0.79  \pm  0.01$   &    $54.7  \pm  0.4$   &   $0.76  \pm  0.01$   &   $53.3  \pm  0.3$   \\
          $>$58222   &   $0.76  \pm  0.02$   &    $51.4  \pm  0.6$   &   $0.87  \pm  0.02$   &    $50.5  \pm  0.8$   &   $0.86  \pm  0.02$   &  $45.8  \pm  0.6$   \\
        \hline
        \end{tabular}
        \label{tab:maxi_ppa}
\end{table*}

\section{Data}\label{sect:data}

\subsection{Optical polarimetric observations}

Polarimetric observations of MAXI~J1820+070 were performed with the Dipol-2 polarimeter \citep{PBB14} on the remotely controlled Tohoku 60~cm (T60) telescope at Haleakala Observatory, Hawaii. 
To achieve the high accuracy in polarimetric measurements, the Dipol-2 exploits the `double-image' design, which effectively eliminates the errors arising due to variations in seeing and transparency. 
The optical beam from the star is split into two parallel, orthogonally polarized beams (e- and o-rays), giving rise to two stellar images that are recorded simultaneously in two separated parts of each of three ({\it BVR}) CCDs. 
The two orthogonally polarized beams from the sky overlap on each image, and the total sky intensity is recorded in both of them.
The stellar fluxes in both images are then extracted via aperture photometry.
Because the sky intensities are equal for both images, its polarisation is automatically cancelled.
The resulting stellar polarisation is completely free of systematic errors caused by sky polarisation, even if it is variable over the observing run.

The observations were conducted for 12 nights from 2018 March 17 until April 25 (MJD~58195--58234).  
In every observing night, 10 to 48 measurements of Stokes parameters $q$ and $u$ were made simultaneously in the {\it BVR}-filters, and the average nightly polarisation values, $P$, and polarisation angle, PA, were computed. 
The instrumental polarisation was determined from the observations of more than 20 non-polarized nearby stars. 
The magnitude of the instrumental polarisation for the T60 telescope is lower than 0.01\% and therefore negligible in this case. 
The PA zero-point was measured using observations of the highly polarized standards HD25433 and HD161056.  
The observational errors of the normalized Stokes parameters $q$ and $u$ were computed as the standard errors of the weighted mean values.
A detailed description of the observation procedure, calibration, and the data treatment techniques is given in \citet{KBP17}.
The nightly averages of our polarimetric observations of MAXI~J1820+070 are listed in Table~\ref{tab:maxi_ppa} and shown in Fig.~\ref{fig:lc_scale}.

% fig 2
\begin{figure}
\center{\includegraphics[width=0.85\columnwidth]{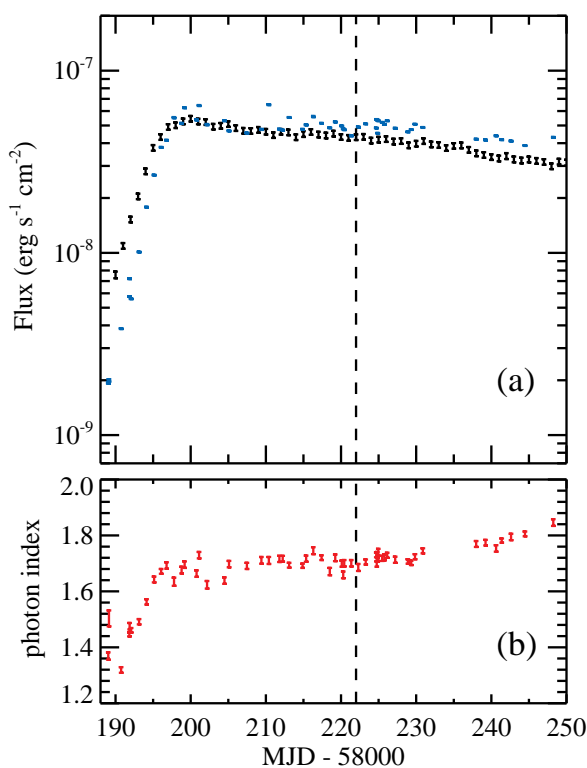}}
    \caption{(a) MAXI~J1820+070 outburst light curves at the initial stages as determined by the {\it Swift}/BAT (black error bars) and the {\it Swift}/XRT (blue error bars) in the 15--50 keV and 0.5--10 keV bands.
    (b) Evolution of the photon index as seen by the {\it Swift}/XRT. The vertical line marks MJD~58222. 
    }
    \label{fig:lc_x}
\end{figure}

\subsection{X-ray data}

A very good coverage of the initial stages of the outburst was
obtained through the {\it Neil Gehrels Swift Observatory}
\citep{2004ApJ...611.1005G}. The source was observed in the soft (0.5--10 keV) and
hard (15--50 keV) X-ray ranges using the XRT telescope \citep{2005SSRv..120..165B} and the BAT monitor
\citep{2005SSRv..120..143B}, respectively. 
Because the source is very bright, all XRT observations were performed in windowed timing (WT) mode. 
The spectrum  was extracted using the online tools
\citep{2009MNRAS.397.1177E}\footnote{\url{http://www.swift.ac.uk/user_objects/}}
provided by the UK Swift Science Data Centre.  
Only zero-grade (single pixel) events were included in the products.

% fig 3
\begin{figure}
\center{\includegraphics[width=0.85\columnwidth]{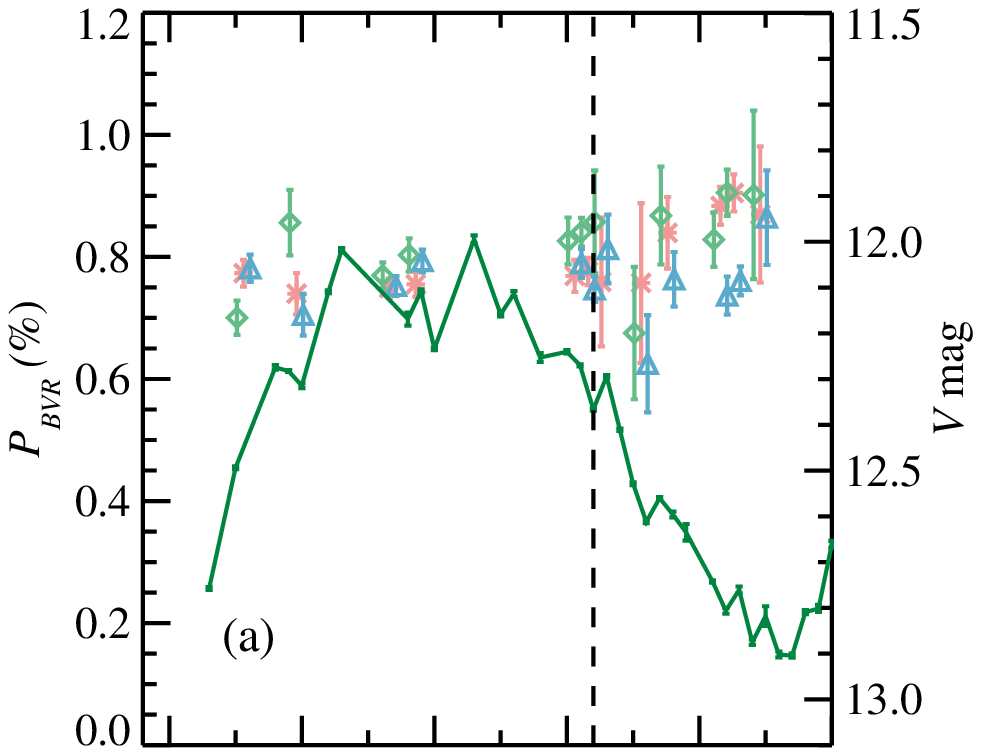} \\ \includegraphics[width=0.85\columnwidth]{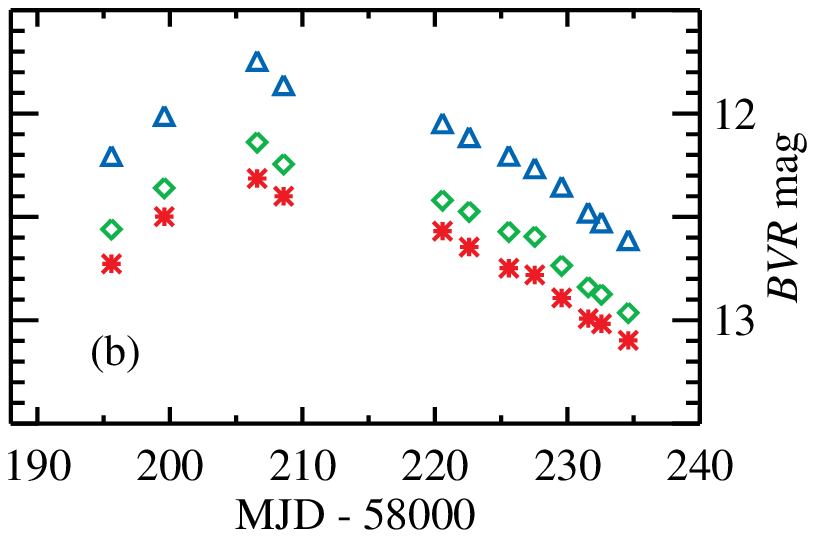}}
    \caption{ (a) Light curve of MAXI~J1820+070 in the $V$ filter (green line with error bars, from AAVSO observations). 
polarisation measurements are overplotted with pale-coloured symbols. As in Fig.~\ref{fig:lc_scale}, the dates  for  the polarisation measurements in the $B$ and $V$ filters are shifted by 0.5~d for clarity. The vertical line marks MJD~58222. 
(b) Evolution of the Dipol-2 $B$ (blue triangles), $V$ (green diamonds) and $R$ (red crosses) magnitudes, calculated from the relative flux of the target and field star, hence the the zero-point is uncertain.}
    \label{fig:lc_xv}
\end{figure}

The source flux in each XRT observation was determined based on the results of spectral fitting with the {\sc xspec} package assuming the power-law model modified by the photoelectric absorption (model {\sc phabs*po}).  
Because of the known calibration uncertainties at low energies,\footnote{\url{http://www.swift.ac.uk/analysis/xrt/digest_cal.php}}
we restricted our spectral analysis to the 0.5--10 keV band.  
In the hard X-ray band (15--50 keV) we used results from the BAT transient
monitor\footnote{\url{http://swift.gsfc.nasa.gov/results/transients/index.html}}
\citep{2013ApJS..209...14K}. 
Conversion of the {\it Swift}/BAT count rate into flux units was performed assuming Crab-like spectrum of the source and the count rate from the Crab Nebula of 0.22 counts s$^{-1}$ cm$^{-2}$.  
The final absorption-corrected light curve of MAXI~J1820+070 is shown in Fig.~\ref{fig:lc_x}(a) with blue and black points for XRT and BAT fluxes, respectively. 
Fig.~\ref{fig:lc_x}(b) presents the corresponding evolution of the photon index extracted from the XRT data.
We compared the measured fluxes and hardness ratios to those obtained from the MAXI monitor and found a good agreement.
However, the MAXI data have a large gap close in time to our observations, therefore we did not use them further in our analysis.

% fig 4
\begin{figure}
\center{\includegraphics[width=0.85\columnwidth]{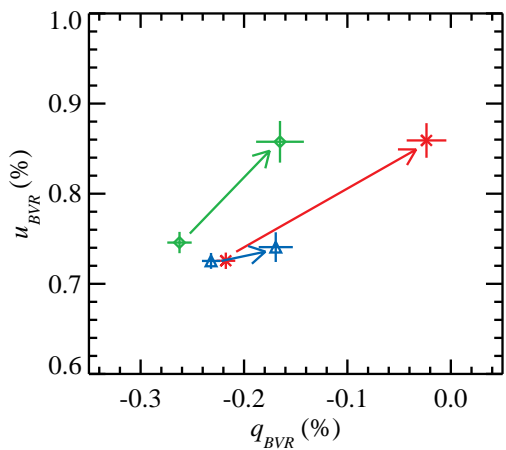}}
\caption{Change in observed polarisation in MAXI~J1820+070. 
Symbols denote weighted-average polarisation before and after MJD~58222 in different filters: $B$ (blue triangles), $V$ (green diamonds), and $R$ (red crosses).
Arrows show the direction of the polarisation change.
The vector of additional polarisation in the $R$ filter is about 50\% longer than that in the $V$ filter. 
    }
    \label{fig:qu_change}
\end{figure}

\subsection{Optical photometry}

We used the public AAVSO {\it V}-band light curves.
For each day we calculated the averaged magnitude and the standard error of the mean, which we report as the magnitude error (see the dark green line in Fig.~\ref{fig:lc_xv}a).
We also extracted the {\it BVR} relative fluxes of MAXI~J1820+070 and a close field star taken during our polarisation measurements (see Fig.~\ref{fig:lc_xv}b).
Although these photometric measurements suffer from the uncertainties related to possible changes in the flux of the comparison star, the {\it V}-band light curve is found to be in good agreement with the AAVSO light curve.
We additionally retrieved the {\it Swift}/UVOT {\it U}-filter light curves, but in most of these observations, MAXI~J1820+070 shows signs of saturation, and thus we did not consider these data in the paper.

\section{Results}\label{sect:results}

\subsection{Evolution of the observed polarisation}

The light curves of polarisation degree and PA in three filters are shown in Fig.~\ref{fig:lc_scale}.
A small but significant polarisation at a level of $\sim0.7$\% is detected in the source direction, with a hint of an increase towards the end of observations.
The PA initially remains the same within the errors, but after MJD~58222, there is a clear trend of its decrease.

In order to understand the nature of the observed variability, we compared the variation in polarisation with photometric $V$-band and X-ray light curves. 
The observed total flux of $\sim$$10^{-7}$~erg~s$^{-1}$~cm$^{-2}$ in the 0.3--10 and 15--50~keV bands (see Fig.~\ref{fig:lc_x}a) and the distance estimate from the {\it Gaia} DR2 data ($\sim$$4$~kpc, \citealt{GRJ18}) allowed us to roughly estimate the luminosity at the peak of the outburst $L_{\rm X}$$\sim$$2 \times 10^{38}$~erg~s$^{-1}$. 
In spite of this rather high luminosity, the source remained in the hard/low state, as evidenced by the measured X-ray spectral slope (Fig.~\ref{fig:lc_x}b).\footnote{The transition to the soft state occurred at the beginning of July 2018 \citep{ATel11820,ATel11823}.}
The spectral index was stable before MJD~58222, but started to increase after this.
Furthermore, we see a substantial drop of {\it V}-band flux around the same date (from AAVSO data, Fig.~\ref{fig:lc_xv}), when we detect the departure of the PA from $\sim55\degr$.
We find a similar decrease in all filters of our {\it BVR} photometric light curves.
Conversely, the X-ray flux was decaying smoothly around that date, suggesting a decoupling of the X-ray and optical fluxes 
\citep[as also reported by][]{ATel11574}.

To check the dependence of the polarisation properties on time, we split the observations into two parts, separated by MJD~58222.
We calculated the average of these parts (see Table~\ref{tab:maxi_ppa}) and plot them in Fig.~\ref{fig:qu_change} in the Stokes parameter plane.
The vectors of an additional polarisation component in the $V$ and $R$ bands  are nearly parallel. 
This suggests that the evolution is caused by an emerging polarized component.

The evolution is apparent in all bands. 
The statistical significance of change in $q$ and $u$ Stokes parameters can be estimated using the multivariate Hotelling $T^2$ test \citep{Hotelling1931}.  
The values for $t^2$ are 12.8, 35.9, and 150 for the $B$,  $V,$ and $R$ filters, respectively. 
The sizes $n = n_1 + n_2$ of the corresponding data sets are 398, 413, and 415, where $n_1$ and $n_2$ are the number of observations used to calculate average polarisation before and after MJD~58222, respectively. 
The corresponding values of the variable $f=(n-3)t^2/[2(n-2)]$ (which follows an $F$-distribution with parameters 2 and $n-3$) are 6.4, 17.9, and 74.7, which give the probabilities that polarisation has not changed  $2\times 10^{-3}$ (i.e. above  $3\sigma$),  $3.5\times 10^{-8}$ , and $2\times 10^{-28}$ 
\citep[see][for details of applying the Hotelling $T^2$ test to the polarisation data]{KBP17}. 
To verify that the polarisation evolution is not caused by instrumental effects, we computed the average polarisation of a field star (star 1 in Table~\ref{tab:field_ppa}) before and after MJD~58222 and obtained that the polarisation degrees agree well within errors.

\begin{figure*}
\center{\includegraphics[width=0.7\columnwidth]{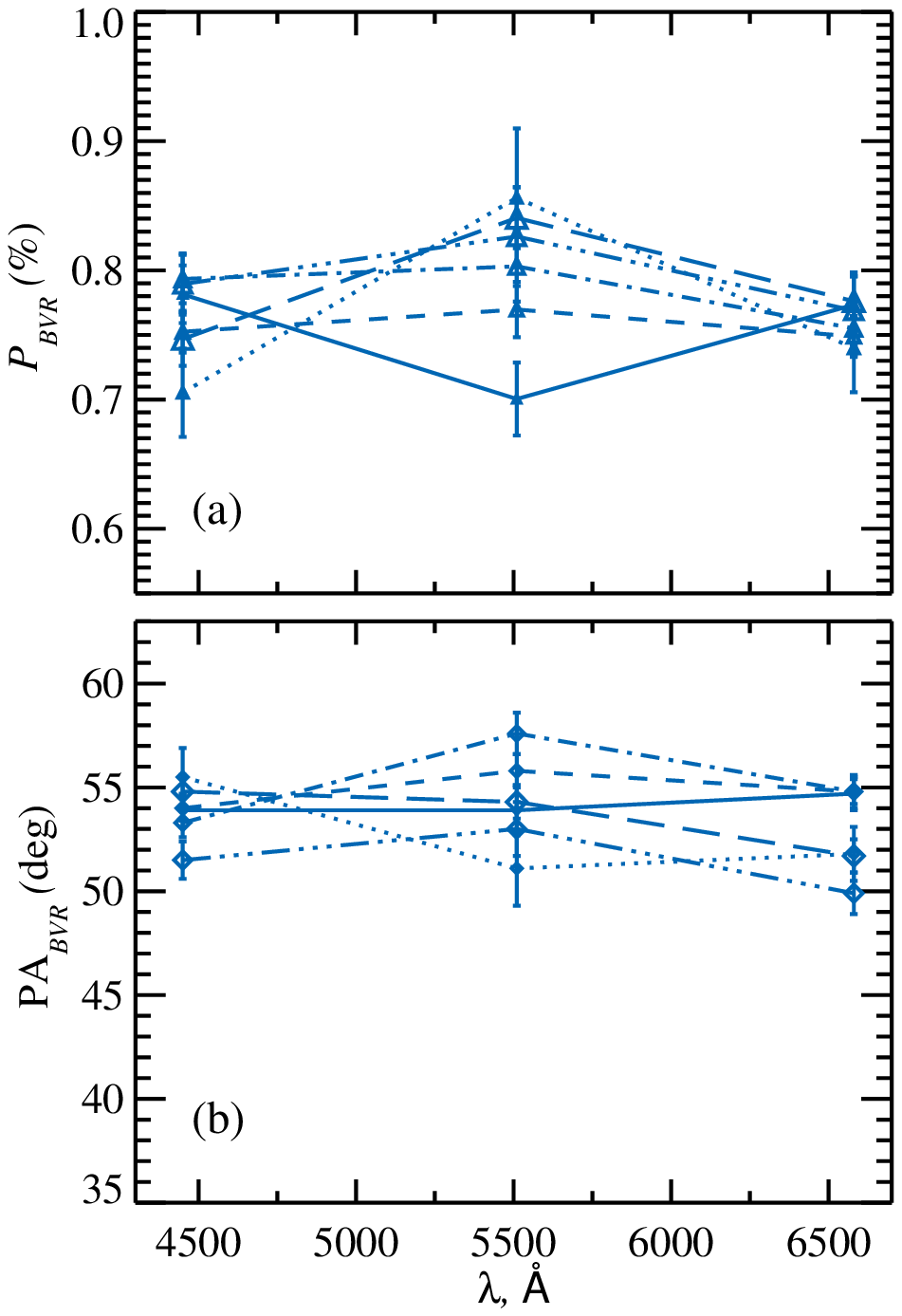}
\includegraphics[width=0.7\columnwidth]{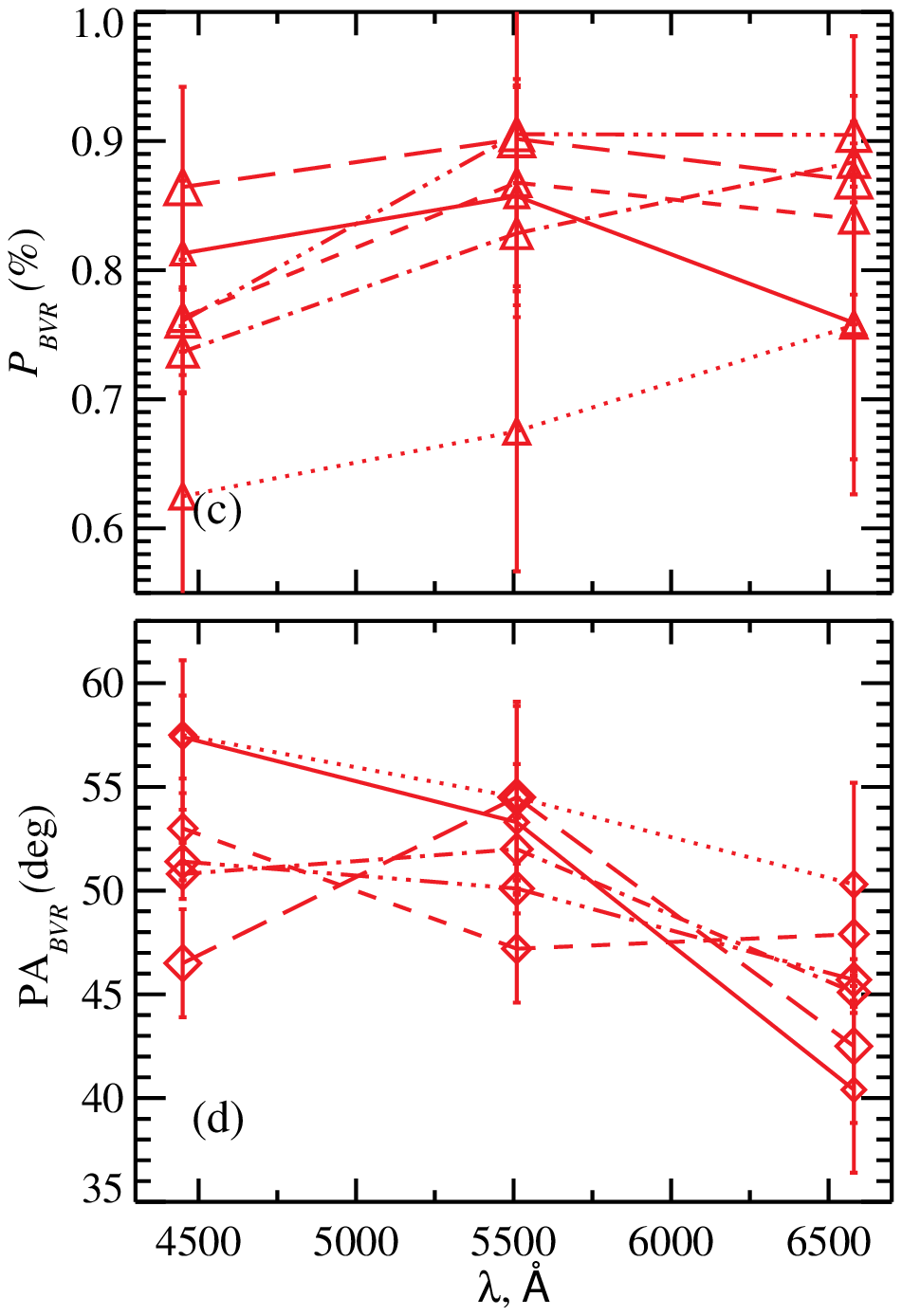}}
\caption{Spectral evolution of the polarisation degree and the PA before MJD~58222 (a, b) and after MJD~58222 (c, d). Line styles in the order of increasing date: solid, dotted, dashed, dot-dashed, triple-dot-dashed, long-dashed. 
Both $P$ and PA show almost flat spectral dependence in panels a and b. The weak dependence of both $P$ and PA on the wavelength is visible in panels c and d.} 
\label{fig:spec_obs}
\end{figure*}

The spectral dependence of polarisation degree and PA of the source before and after MJD~58222 are shown in Fig.~\ref{fig:spec_obs}.
Each line corresponds to the measurement averaged over the night. 
The spectral dependence of polarisation degree and PA before MJD~58222 is almost flat, in contrast to that after MJD~58222, which shows a clear skew towards the $R$ band.  

% fig 5

% fig 6
\begin{figure}
\center{\includegraphics[width=0.85\columnwidth]{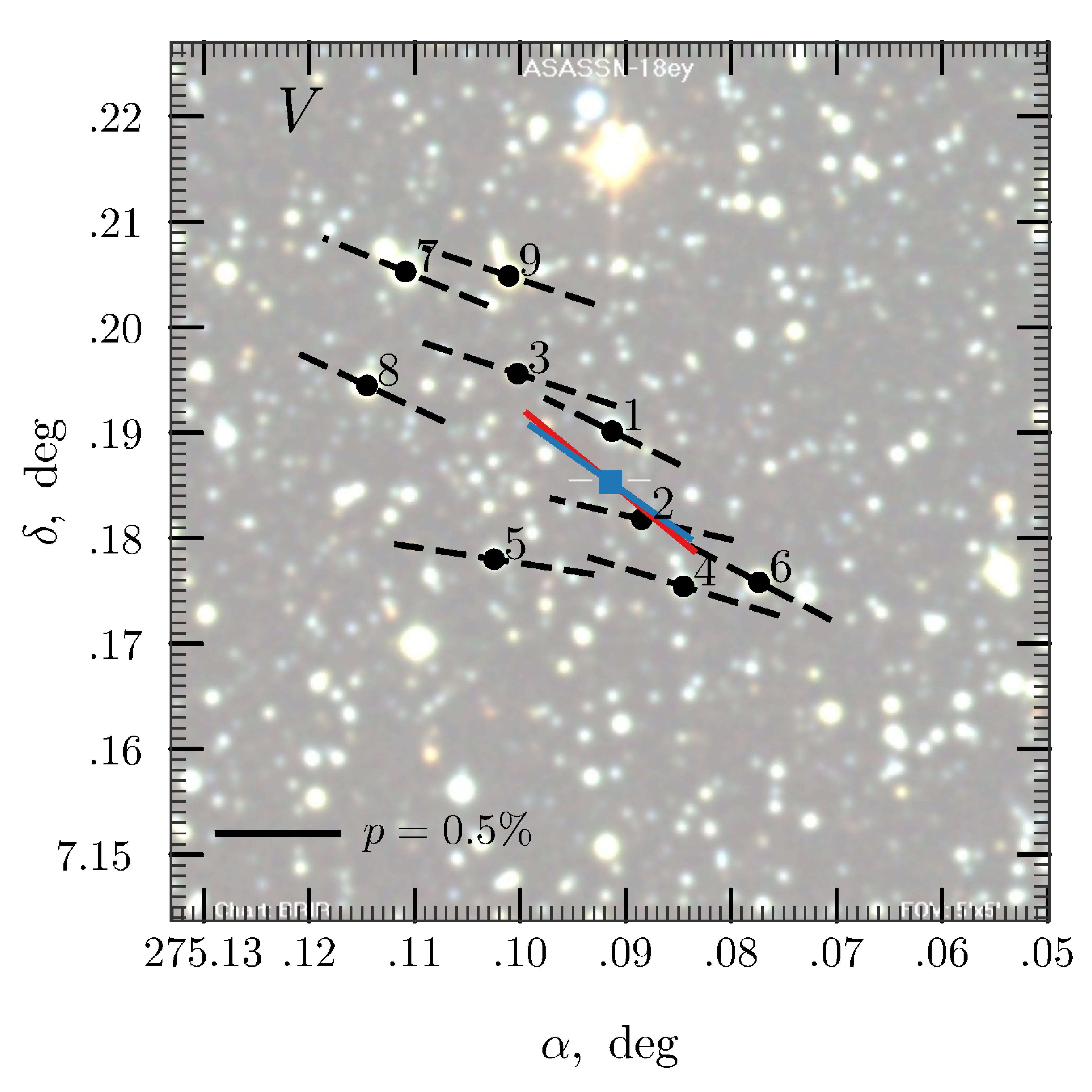}}
\caption{polarisation vectors of the field stars and MAXI~J1820+070 in the $V$ filter. The background image is taken from \citet{ATel11400}. The blue square denotes the position of MAXI~J1820+070, for which two polarisation vectors are shown before and after MJD~58222 with blue and red lines. 
    }
    \label{fig:field}
\end{figure}

% fig 7 
\begin{figure}
\center{\includegraphics[width=0.75\columnwidth]{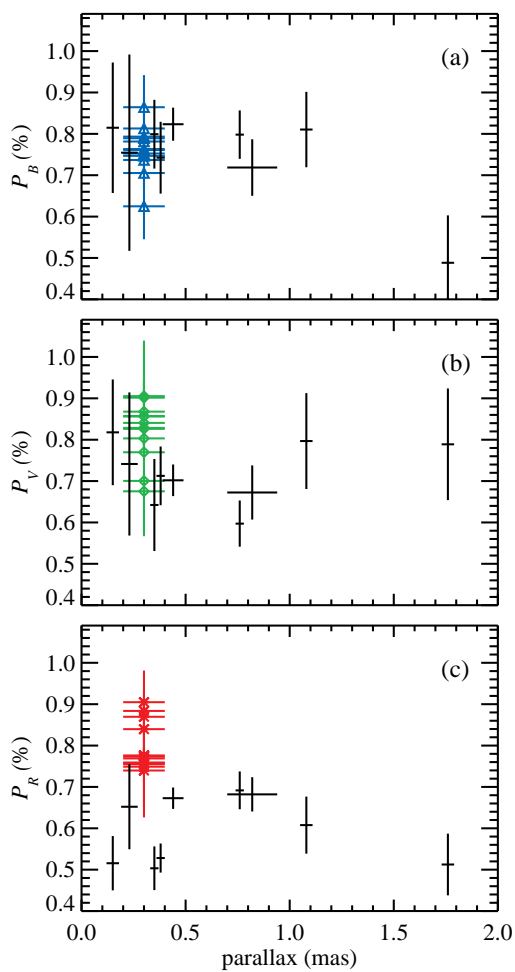}}
    \caption{polarisation degree as a function of parallax for MAXI~J1820+070 (coloured symbols) and the field stars (black crosses).}
    \label{fig:pol_plx_scale}
\end{figure}

% fig 8
\begin{figure}
\center{\includegraphics[width=0.75\columnwidth]{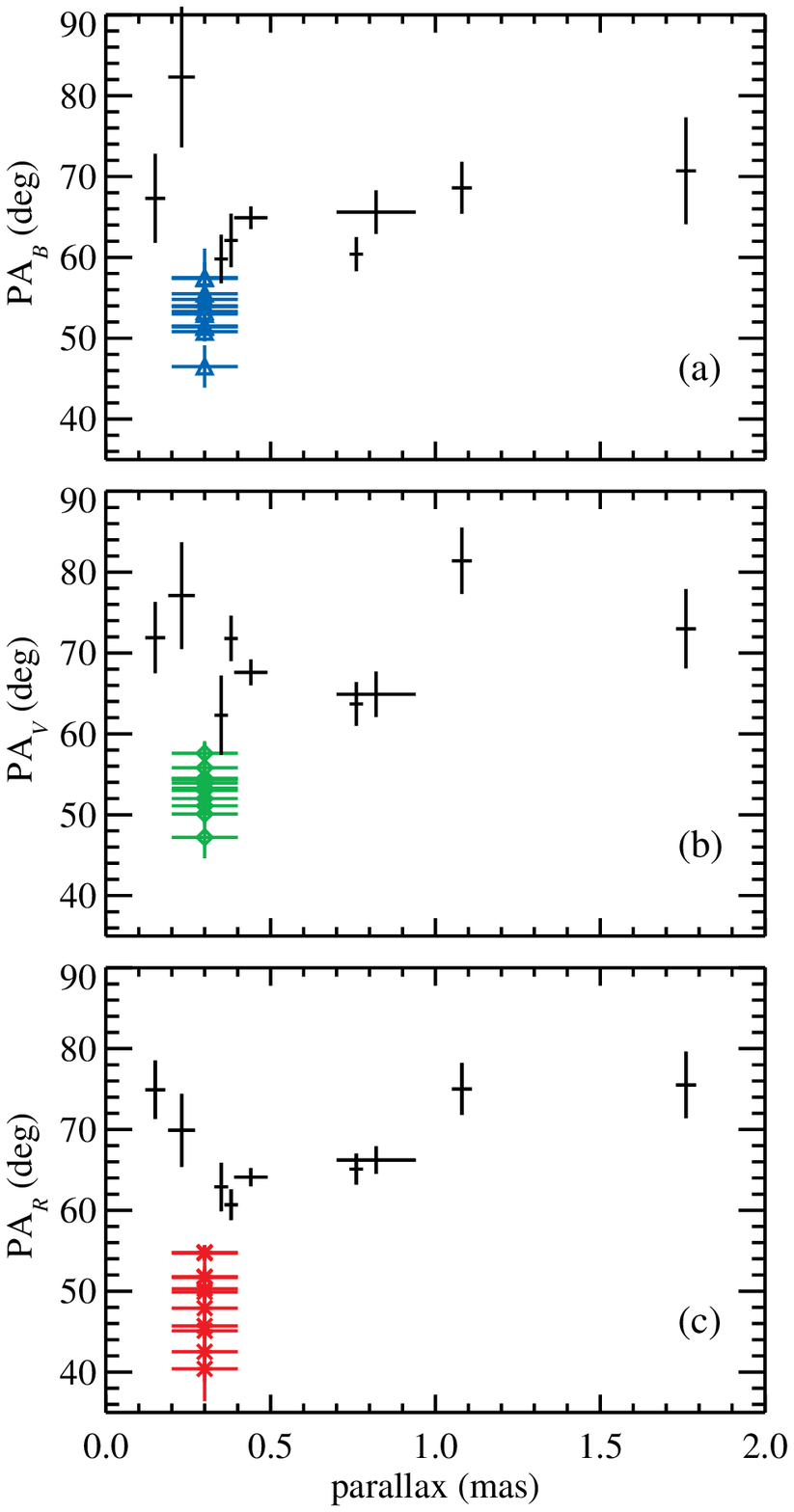}}
    \caption{Same as in Fig.~\ref{fig:pol_plx_scale}, but for the polarisation angle. }
    \label{fig:pa_plx_scale}
\end{figure}

\subsection{Interstellar polarisation}

The upper limit for the polarisation of the interstellar medium (ISM) in the direction of MAXI~J1820+070 can be estimated from the reddening of the source \citep{Serk75}. 
The colour excess in the direction of MAXI~J1820+070 is only $E(B-V)=0.19$~mag as given by the IRSA database, while the estimate from the hydrogen column density gives $E(B-V)=0.163\pm0.007$ \citep{ATel11418}.
Hence, the expected ISM polarisation degree is $P_{\rm ISM}<9\times E(B-V)=1.71$\% or $1.46$\% for the two estimates.

To more accurately estimate the contribution from the ISM polarisation to that observed from MAXI~J1820+070,  we studied polarisation of the field stars. 
This method of estimating the ISM polarisation in the source direction has been shown to give sufficiently accurate results \citep{KBP17}.
The resulting measurements for nearby field stars are given in Table~\ref{tab:field_ppa}.
The {\it V}-filter sky map of the field around MAXI~J1820+070, together with the polarisation vectors of the field stars and the average values of the source before and after MJD~58222, are shown in Fig.~\ref{fig:field}.

% Table 2 field stars
\begin{table*}
\centering 
\caption{Polarimetric data for the field stars and their weighted average. 
}
\begin{tabular}{cccccccc}
                \hline
        &       & \multicolumn{2}{c}{$B$} & \multicolumn{2}{c}{$V$} & \multicolumn{2}{c}{$R$} \\
 Star & $\pi$~(mas)     &       $P$ (\%) & PA ($\degr$) & $P$ (\%) & PA ($\degr$) & $P$ (\%) & PA ($\degr$)\\

                \hline
        1   &   $0.76  \pm  0.02$   &   $0.80  \pm  0.06$   &    $60.4  \pm  2.1$   &   $0.60  \pm  0.06$   &    $63.7  \pm  2.7$   &   $0.69  \pm  0.05$   &   $65.1  \pm  1.9$   \\
        2   &   $0.23  \pm  0.04$   &   $0.75  \pm  0.24$   &   $82.3  \pm  8.7$   &   $0.74  \pm  0.17$   &   $77.1  \pm  6.6$   &   $0.65  \pm  0.10$   &   $69.9  \pm  4.5$   \\
        3   &   $0.15  \pm  0.03$   &   $0.81  \pm  0.16$   &    $67.3  \pm  5.5$   &   $0.82  \pm  0.13$   &    $71.9  \pm  4.4$   &   $0.52  \pm  0.07$   &   $74.9  \pm  3.6$   \\
        4   &   $1.76  \pm  0.03$   &   $0.49  \pm  0.11$   &    $70.7  \pm  6.6$   &   $0.79  \pm  0.13$   &    $73.0  \pm  4.9$   &   $0.51  \pm  0.07$   &   $75.5  \pm  4.1$   \\
        5   &   $1.08  \pm  0.03$   &   $0.81  \pm  0.09$   &    $68.6  \pm  3.2$   &   $0.80  \pm  0.12$   &    $81.4  \pm  4.1$   &   $0.61  \pm  0.07$   &   $75.0  \pm  3.2$   \\
        6   &   $0.35  \pm  0.02$   &   $0.80  \pm  0.08$   &    $59.8  \pm 3.0$   &   $0.64  \pm  0.11$   &    $62.3  \pm  4.9$   &   $0.50  \pm  0.05$   &   $62.9  \pm  3.0$   \\
        7   &   $0.44  \pm  0.05$   &   $0.82  \pm  0.04$   &    $64.9  \pm  1.4$   &   $0.70  \pm  0.04$   &    $67.6  \pm  1.6$   &   $0.67  \pm  0.03$   &   $64.1  \pm  1.1$   \\
        8   &   $0.82  \pm  0.12$   &   $0.72  \pm  0.07$   &    $65.6  \pm  2.7$   &   $0.67  \pm  0.07$   &    $64.9  \pm  2.8$   &   $0.68  \pm  0.04$   &   $66.2  \pm  1.7$   \\
        9   &   $0.38  \pm  0.02$   &   $0.74  \pm  0.09$   &    $62.1  \pm  3.3$   &   $0.71  \pm  0.07$   &    $71.8  \pm  2.8$   &  $0.53  \pm  0.04$   &   $60.7  \pm  1.9$   \\
         \hline
2,3,6,7,9   &&                $0.80  \pm   0.03$   &    $64.2    \pm   1.2$   &   $0.70  \pm   0.03$  &     $68.6    \pm   1.3$   &   $0.60  \pm   0.02$    &   $64.1 \pm  0.9$ \\   
      \hline
        \end{tabular}
        \label{tab:field_ppa}
\end{table*}

We compared the observed polarisation degree and PA of MAXI~J1820+070 with those of the field stars.
All stars have {\it Gaia}~DR2 counterparts, allowing a comparison of their polarisation properties as a function of distance (parallax).
The polarisation degree and PA of the field stars are almost independent of distance (see Figs~\ref{fig:pol_plx_scale} and \ref{fig:pa_plx_scale}).
The source $B-$ and $V$ -band polarisation degrees are roughly consistent with those of the field stars, but the $R$ band polarisation is significantly different from that of the field stars.
The polarisation angles substantially deviate from the nearby field stars in all bands.

\begin{figure}
\center{\includegraphics[width=0.75\columnwidth]{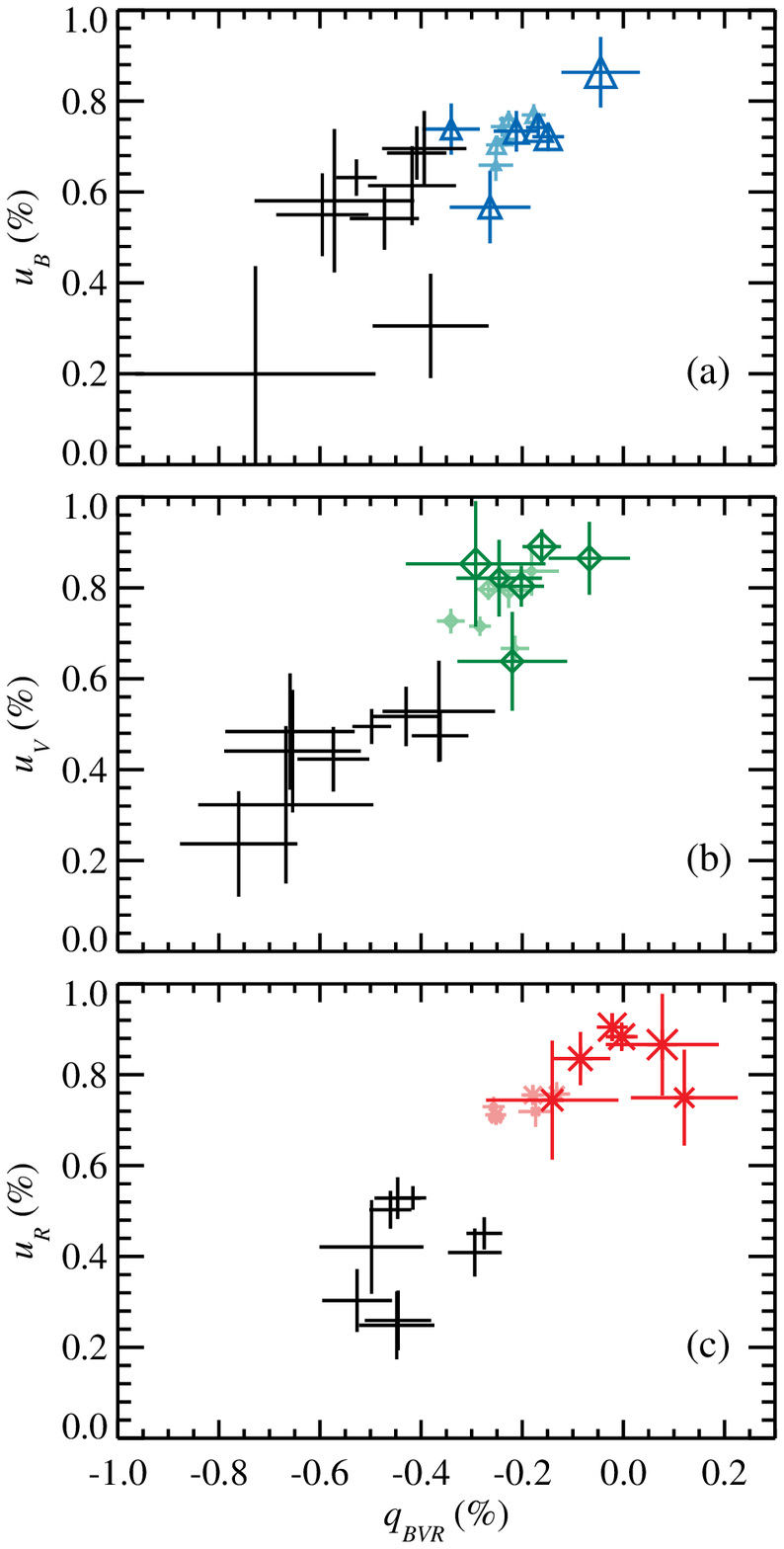}}
\caption{Normalized Stokes parameters for MAXI~J1820+070 (coloured symbols) and the field stars (black crosses) in the $q-u$ plane. 
The pale and the saturated colours correspond to observations of MAXI~J1820+070 made before and after MJD~58222, respectively. 
The size of the symbols grows with MJD. }
    \label{fig:qu}
\end{figure}

In Fig.~\ref{fig:qu} we plot the normalized Stokes parameters of MAXI~J1820+070 and the field stars.
In all three filters, the Stokes vectors of the source cluster in a region separate from that occupied by the field stars in the $q-u$ plane.
The difference is most prominent in the $R$ band, where the MAXI~J1820+070 points are entirely isolated.
This points towards substantial intrinsic polarisation of MAXI~J1820+070 throughout all observations.

% fig 9

% fig 10
\begin{figure}
\center{\includegraphics[width=0.75\columnwidth]{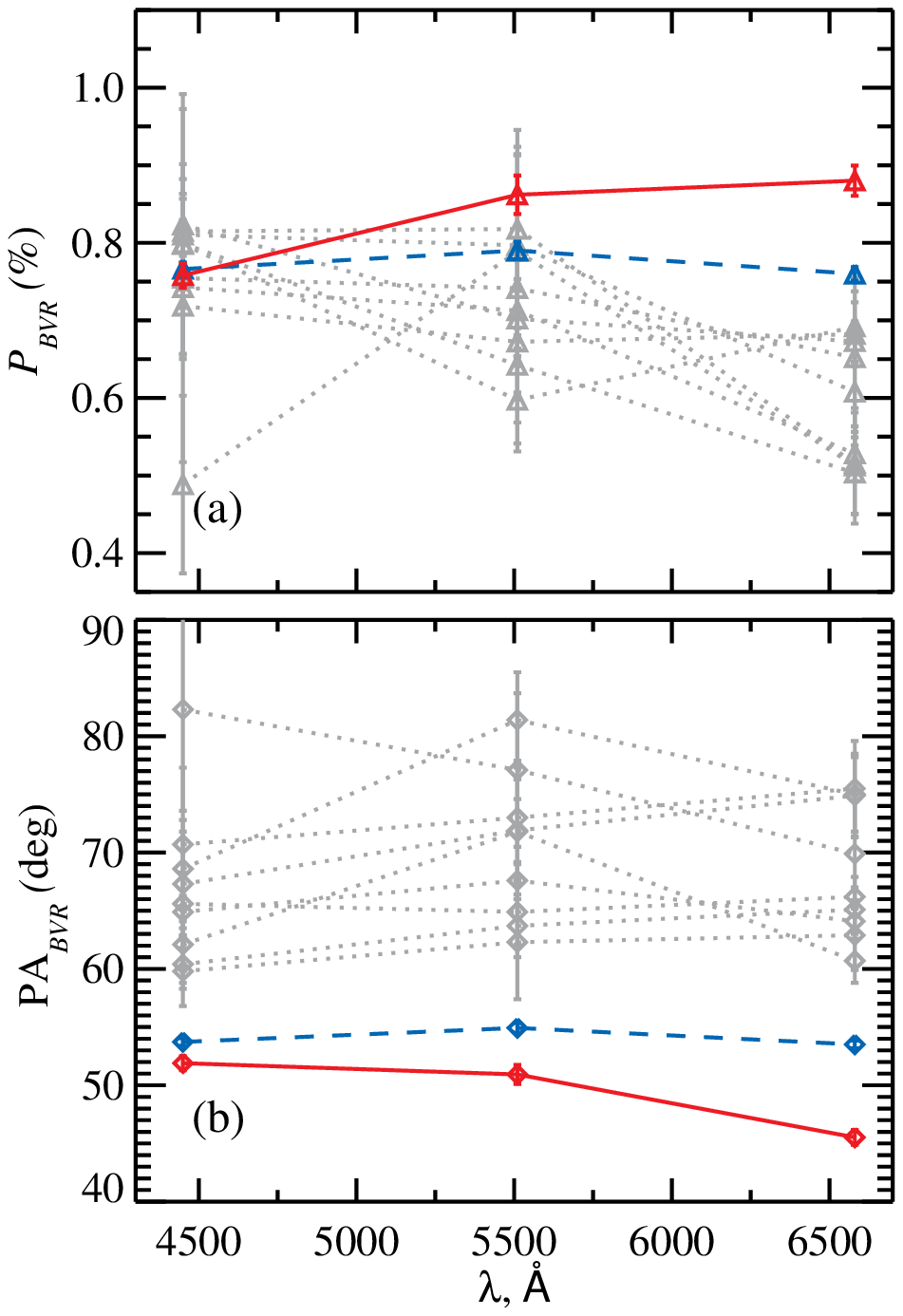}}
\caption{Wavelength dependence of the polarisation degree and the PA of the field stars (grey dotted lines) and of the weighted averages for MAXI~J1820+070 (blue dashed and red solid lines for observations before and after MJD~58222, respectively).}
    \label{fig:spec_field}
\end{figure}

To better emphasise this, we plot the average {\it P} and PA in Fig.~\ref{fig:spec_field}: the blue dashed lines correspond to the average for observations made before MJD~58222, and the red solid line corresponds to the average computed after that date.
The spectral distributions of field stars are shown with the grey dotted lines.
We note that the flat polarisation degree and PA spectral distribution of the source is very similar to that of the field stars before MJD~58222.
However, the absolute value of the source PA is different from those of the field stars in all observations. 
Furthermore, the spectral dependence of the polarisation degree observed after MJD~58222 is skewed towards $R$ and has a strong dependence of PA on wavelength. 
This is in contrast to that observed in the field stars.
This further supports the suggestion that there is a substantial polarisation of the source with an origin other than interstellar.

We estimated the contribution of the ISM polarisation to that of the source in two ways.
We first considered a star that was closest in angular separation from MAXI~J1820+070 and was reasonably similar in parallax and used its polarimetric characteristics as a proxy for ISM polarisation. 
In our case, this is star 2 from Table~\ref{tab:field_ppa}. 
As an alternative, we considered a set of stars close to the source and averaged their polarimetric measurements. 
For this we chose five stars from Table~\ref{tab:field_ppa}, whose parallaxes were most similar (0.15--0.5 mas) to that of the source, and calculated the weighted average of their Stokes vectors, with the weights being inversely proportional to the square of individual errors. 
The results are given in the bottom line of Table~\ref{tab:field_ppa}. 
The largest difference (about 2--3$\sigma$) between these average values and those for star 2 are in the PA of the $B$ and $V$ filters.

\subsection{Intrinsic polarisation}

% fig 11
\begin{figure}
\center{\includegraphics[width=0.75\columnwidth]{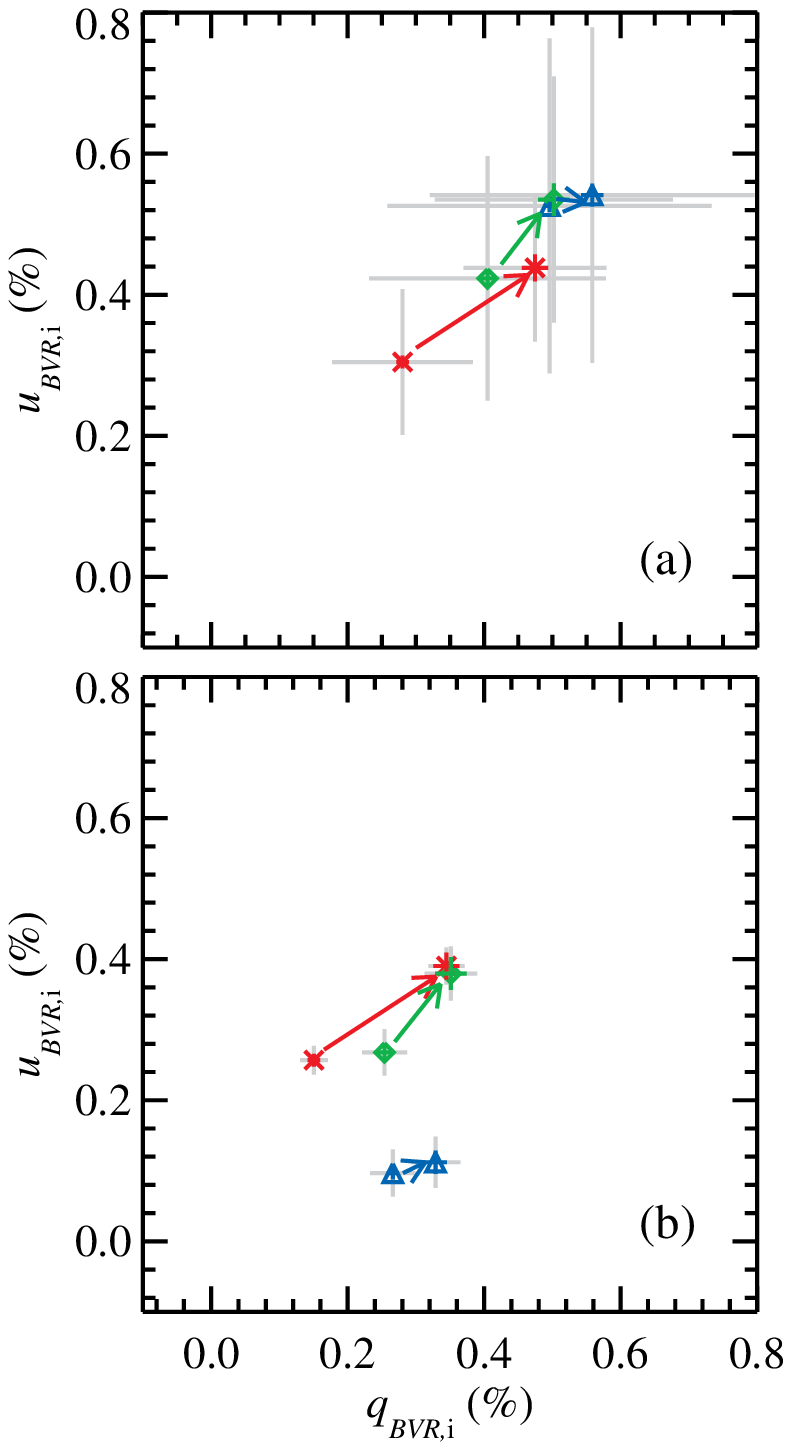}}
    \caption{Change in intrinsic Stokes parameters of MAXI~J1820+070. The weighted average polarisation is shown before and after MJD~58222 in three filters: $B$ (blue triangles), $V$ (green diamonds), and $R$ (red crosses).
    The two panels correspond to two different estimates for the ISM polarisation: (a) from the closest field star 2  (see Table~\ref{tab:field_ppa}),  and (b) from the five-star (2, 3, 6, 7, and 9) average.
    Coloured error bars correspond to the error of the mean of the source averages, and grey error bars also include the  uncertainties in the estimate of ISM polarisation.  
    }
    \label{fig:qu_intr}
\end{figure}

% Table 3 intrinsic polarisation
\begin{table*}
        \centering 
        \caption{Intrinsic polarisation of MAXI~J1820+070.  The statistical errors are given after the average values, 
         and the systematic uncertainties arising from the subtraction of the ISM polarisation are given in brackets.}
        \begin{tabular}{cccccccc}
                \hline
 & & \multicolumn{2}{c}{$B$} & \multicolumn{2}{c}{$V$} & \multicolumn{2}{c}{$R$} \\
$P_{\rm ISM}$  & MJD &$P$ (\%) & PA ($\degr$) & $P$ (\%) & PA ($\degr$) & $P$ (\%) & PA ($\degr$)\\
      \hline
 Star 2  &  $<$58222    &  $0.72  \pm  0.01(.24)$    &    $23.4  \pm  0.4(9.4)$    &   $0.59  \pm   0.01(.17)$    &    $23.1  \pm  0.5(8.5)$    &   $0.41  \pm   0.01(.10)$    &   $23.7  \pm  0.7(7.1)$ \\
            &  $>$58222     &  $0.78  \pm  0.02(.24)$    &    $22.1  \pm  0.7(8.8)$    &   $0.73  \pm   0.02(.17)$    &    $23.4  \pm  0.8(6.8)$    &   $0.65  \pm   0.02(.10)$    &   $21.4  \pm  0.9(4.6)$ \\
      \hline 
2,3,6,7,9  & $<$58222  &  $0.28  \pm  0.01(.03)$    &    $10.0  \pm  1.0(3.4)$       &    $0.37  \pm  0.01(.03)$    &    $23.3  \pm  0.8(2.6)$    &    $0.30  \pm  0.01(.02)$    &    $29.8  \pm  0.9(2.0)$ \\
                &  $>$58222 &  $0.35  \pm  0.02(.04)$    &    $ 9.4  \pm   1.6(3.0)$    &    $0.52  \pm  0.02(.04)$    &    $23.6  \pm  1.1(2.1)$    &    $0.52  \pm  0.02(.03)$    &    $24.3  \pm  1.1(1.5)$ \\
      \hline
        \end{tabular} 
\begin{flushleft}{ 
{\it Notes.} The intrinsic polarisation was obtained using two estimates of the ISM polarisation: from the closest field star 2 (two top rows), and from the five-star average (two bottom rows). Upper and lower rows correspond to the average polarisation parameters before and after MJD~58222, respectively.
}\end{flushleft} 
        \label{tab:intrinsic_p}
\end{table*}

Using the estimates for the interstellar polarisation, we computed the intrinsic polarisation of the source.
From the observed Stokes averages of the source we subtracted the field star sample mean and show the results in Fig.~\ref{fig:qu_intr} for two cases: the $P_{\rm ISM}$ estimate using the closest field star 2 and using the five-star average.
The resulting estimates are given in Table~\ref{tab:intrinsic_p}.

For the five-star ISM polarisation estimate (Fig.~\ref{fig:qu_intr}b), we find that the angle of intrinsic polarisation is the same (within errors) for the $V$ and $R$ bands, and it is different from that in the $B$ band.
The polarisation vector changes in the $q-u$ plane are almost the same in both filters, but the Stokes vectors before and after MJD~58222 are not strictly parallel, implying a possible evolution of the  polarisation degree and angle between the observations.

When we estimate the interstellar contribution using only the closest star 2, we find that the intrinsic PA of the source is consistent in all three photometric filters and for all dates, that is, all data points lie very close to one line with the start in the origin.
This result is striking, given the large error bars, which mostly come from the star 2 polarisation uncertainties.
This implies that the PA did not change, but rather the intrinsic polarisation degree of the source has grown  after MJD~58222.

% fig 12
\begin{figure}
\center{\includegraphics[width=0.75\columnwidth]{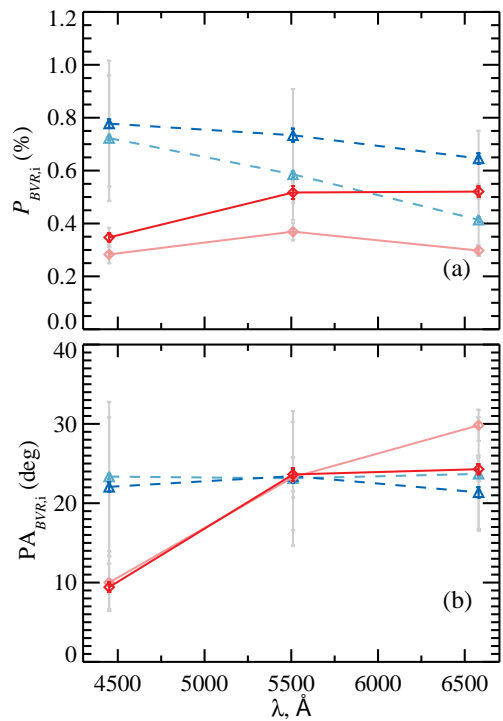}}
    \caption{Wavelength dependence of the intrinsic (a) polarisation degree and (b) PA. Weighted averages are shown with the pale and saturated colours for observations before and after MJD~58222, respectively. The blue dashed and red solid lines correspond to the intrinsic polarisation calculated using star 2 or the five-star average, respectively, to estimate the ISM polarisation. The values are given in Table~\ref{tab:intrinsic_p}.
Coloured error bars correspond to the error of the source averages, and grey error bars also include the uncertainties of the ISM polarisation estimates. }
    \label{fig:spec_intr}
\end{figure}

The spectral dependence of intrinsic polarisation for two cases of $P_{\rm ISM}$ estimation are shown in Fig.~\ref{fig:spec_intr}.
For the $P_{\rm ISM}$ estimate using star 2, the polarisation degree peaks in $B$, while the polarisation angle is  wavelength independent.
For the five-star $P_{\rm ISM}$ estimate, the PA of the intrinsic component is wavelength dependent.
The spectral dependence of the polarisation degree is almost flat before MJD~58222, but after this date, the polarisation clearly increases towards $R$ band. 
In both cases, the maximal increase of polarisation is observed in $R$, some enhancement is also seen in $V$, and a marginally significant increase is seen in the $B$ filter.

\section{Discussion}\label{sect:discuss}

\subsection{Principal sources of optical polarisation in accreting black holes}

During the outburst of accreting black holes in low-mass X-ray binaries,
the optical luminosity is dominated by the accretion process and the contribution of the secondary star is negligible.
There are, however, many spectral components that can contribute to the optical band, and they can also be polarized to some extent \citep[see the discussion in][]{VPV13,KBP17}. 
In the soft state, the outer parts of the standard accretion disc irradiated by the X-ray emission from the compact object likely dominate the optical flux \citep[see e.g.][]{PV14}. 
In the hard state, there could be at least two additional optical components: the radio-emitting jet \citep[see the review by][]{Fender14SSRv} and the inner hot flow that is believed to dominate in the X-rays \citep{VPV13,PV14,PVR14}.
All these components can be polarized. 

The polarisation degree from the accretion disc depends on the temperature structure of the atmosphere and on the dominating source of the opacity. 
When the electron scattering dominates, the intrinsic polarisation is $<12$\% in the optically thick case and is parallel to the  disc plane, that is, perpendicular to the disc normal \citep{Cha47,Cha60,Sob49,Sob63}.
In the optically thin case,  polarisation is  parallel to the  disc normal and can reach higher values \citep{ST85,BP99,VP04}. 
The opacity by true absorption can significantly affect the polarisation properties. 
It reduces the polarisation degree and leads to a rotation of the polarisation plane by $90\degr$ (relative to the case of pure scattering) at angles close to the disc normal, known as the Nagirner effect \citep{Nagirner62,DGS95,Gnedin97}.
A strong wavelength dependence of the absorption opacity may result in a wavelength-dependent polarisation degree.
This is observed, for example, in Be stars  \citep{Poeckert1978,Poeckert1979} as jumps in polarisation at the edges of the Balmer and Paschen series as well as in the corresponding lines. 

Moreover, the discs in X-ray binaries are likely flared and/or tilted, hence the disc atmosphere local normal changes with distance to the central source.
This may lead to a smooth spectral dependence of polarisation angle because different parts of the disc dominate emission at different wavelengths.

The jets detected in the hard states can become a source of polarised light that is variable in time \citep[e.g.][]{Zdziarski14} because of the nature of synchrotron radiation in the ordered magnetic field. 
The polarisation degree strongly depends on the magnetic field geometry in the emitting region. 
For a power-law distribution of electrons, the maximum polarisation in the optically thin regime (expended in the optical) can reach 70--75\%, depending on the slope of the distribution \citep{RL79}.
For a completely disordered field, the polarisation drops to zero. 

The non-thermal electrons accelerated within the hot inner accretion flow may also emit synchrotron radiation, which can in principle be polarized if the magnetic field structure has an ordered component \citep{VPV13,PV14}.
However, the resulting optical polarisation in this model is expected to be low because the magnetic field in the hot flow is expected to be rather tangled by the turbulent motions within the accretion flow. 
Furthermore, in the case of Galactic black holes, Faraday rotation can significantly affect the observed polarisation produced by both accretion flow and the jet. 
The rotation angle is $\chi_{\rm F} \approx 4\times 10^5 \tau_{\rm T} B_{||,6} \lambda_{\mum}^{2}$, where $ \tau_{\rm T}$ is the Thomson optical depth along the line of sight, $B_{||,6}$ is the parallel component of the magnetic field in units of $10^6$~G, and $\lambda_{\mum}$ is the wavelength in microns.
Thus if the optical emission is produced in an extended region close to the black hole, the expected polarisation is essentially zero.

Optical polarisation can also be produced by scattering of the accretion disc radiation in the outflows from the accretion disc.
For a mildly relativistic outflow with velocity $v/c$ of some tens of percent, which might be associated with the base of the radio jet, the scattering produces polarisation parallel to the jet axis \citep{Beloborodov98,BP99}. 
The maximum polarisation is reached at inclination $\cos i= v/c$. 
The typical polarisation is 10--20\% \citep{Beloborodov98}, but the observed polarisation will be diluted by the disc radiation.  
This scattered component can spectrally hardly be distinguished from the underlying accretion disc spectrum, unless the latter has sharp features associated with the opacity edges \citep{BP99}. 
For a slow wind, the degree of polarisation depends on the inclination angle of the observer, it grows towards high inclinations and is linearly proportional to the optical depth through the outflow. 
The polarisation plane is expected to be perpendicular to the symmetry axis (jet direction).
polarisation of a few percent can easily be produced. 
If the wind is optically thick, the polarisation depends on the shape of the photosphere and can reach a few percent \citep{DGS95,Gnedin97},

\subsection{Origin of polarized optical emission in MAXI~J1820+070}

The nature of the observed polarisation of MAXI~J1820+070 cannot be decisively associated with one of the components described above because of the intrinsic polarisation degree that is lower than 1\%.
Additional information on its wavelength dependence, its variability, and on the broad-band spectral properties is required to understand the source of the polarized light. 
 
\subsubsection{Irradiated disc} 

Numerous spectroscopic observations show asymmetric lines that are skewed towards red, suggesting that a wind is present at the initial stages of an outburst \citep[before $\sim$MJD 58193,][]{ATel11424,ATel11425,ATel11481}.
Observations on MJD~58197 show that the line profile has switched to a symmetric broad shape, resembling the shapes that are observed in the accretion disc \citep{ATel11481}.
Interestingly, the optical spectrum is rather hard \citep[e.g.][]{ATel11533}, and the simultaneous {\it Swift} UVOT and XRT light curves were shown to correlate \citep[data taken on MJD~58188,][]{ATel11432}, favouring a large contribution of the reprocessed radiation to the optical flux.

On the other hand, the substantial variability in the subsecond timescales \citep{ATel11437} is not in line with the reprocessing scenario.
The observed ULTRACAM \citep{Dhillon07} light curves suggest that the variability has a higher amplitude in {\it r'} than in  {\it g'}: the peaks are higher and the dips are deeper in  {\it r'}. 
This means that at least two components contribute to the optical emission: one of them is likely the irradiated disc, and the other is a jet or a hot flow. 
The analysis of the polarimetric data suggests that one of them is polarized and the other is not. 
The drop in the $V$-flux by 50\% after MJD 58222, which coincides with the increase in polarisation, implies that the unpolarized component has decreased. 

We first consider the possibility that the irradiated disc corresponds to the unpolarized component.
The $V$ band is in the Rayleigh-Jeans part of the disc spectrum  \citep{ATel11533} and therefore depends nearly linearly on the disc temperature.
This implies that the decrease in optical flux should result from a decrease in the corresponding irradiating X-ray flux by a factor of 5--6.
In contrast, there is a very little change in the X-ray flux around MJD 58222 (see Fig.~\ref{fig:lc_x}). 
The irradiated disc therefore cannot be the source of the unpolarized emission, but it can be the source of the stable  polarized component.
The low polarisation degree and both alternatives for spectral dependence of the PA, flat and skewed towards $R$, can be explained in the disc scenario (the latter is due to the change in the disc atmosphere local normal).

\subsubsection{Jet} 

The question then is whether the jet can contribute to the optical band and/or be a source of the polarized emission. 
The peak $V$ flux was about 95 mJy (assuming $A_V=0.5$), while the jet flux at 227 and 343 GHz on \MJD{58220}  was 100 and 120 mJy, respectively \citep{ATel11831}. 
This means that the jet could easily produce most of the observed optical flux. 
Increase in polarisation together with the decrease in the optical flux then implies that the jet optical emission has to be unpolarized. 
On the other hand, if the jet emission is produced somewhat similarly to the emission of jets in blazars, it is expected to be intrinsically highly polarized (\citealt*{Impey91}; \citealt{Wills92,Lister01,Marscher02,Ikejiri11}). 
If the jet were the source of the polarized emission, we would expect the increase of polarisation towards longer wavelength due to soft (red) jet spectrum, which is not observed. 

The standard \citet{BK79} jet model is not capable of explaining the observed mid-IR excess \citep{ATel11533}.
An additional jet component associated with instantaneous injection of relativistic electrons at the base of the jet  and their fast cooling \citep{PC09} may be responsible for this. 
It is natural to expect such a jet to emit polarized synchrotron emission because an injection may occur in a small part of the jet with  some preferential magnetic field  direction.
This additional component with a power-law like spectrum contributes to the near-IR and optical bands, and by extrapolating the power law,  we can estimate that its contribution to the $V$-band flux on \MJD{58195--58196} \citep[see][]{ATel11533} is about 30\%.
Thus, in principle, its evolution may be responsible for the observed drop in the $V$ band.
The requirement that this red component is polarized at a level much below one per cent then implies that the jet magnetic field in the emitting region has to be highly tangled.
It is not clear whether this can be incorporated into the current version of this jet model. 

If, however, the emitting region is situated close to the central source, Faraday rotation may destroy the polarisation and satisfy the observational constraints. 
We thus conclude that the standard jet emitting in the radio and the fast-cooling plasma at the base of the jet can both contribute to the optical flux, but these components have to be unpolarized, implying either a disordered magnetic field or a location close to the black hole where Faraday rotation depolarizes their emission.

\subsubsection{Hot flow} 

The observed behaviour of a varying red component and a stable blue component can be naturally explained in the hot accretion flow scenario \citep{VPV13,PV14,PVR14}. 
In the hard state, we expect a contribution of the synchrotron emission from the inner hot flow, which can reproduce the observed red excess to the irradiated disc spectrum. 
On the transition to the soft state, when the truncation radius of the cold disc decreases, this component should decrease first in the red (where the emission from the outer parts of the hot flow contributes) and later in the blue. 
In the soft state, the optical emission is consistent with the irradiated disc, as was reported for example in XTE J1550--564 \citep{PVR14}. 

It is interesting to note that the X-ray spectrum started to soften at about the same time as the $V$ flux dropped and polarisation increased (Figs~\ref{fig:lc_x} and \ref{fig:lc_xv}). 
Thus, if we interpret the softening as a start of the transition to the soft state, we expect a weakening of the red non-thermal hot flow component \citep{KDT13,PVR14}, as observed. 
Because the hot flow spectrum is likely unpolarized, the drop in flux of this component leads to the increase in polarisation. 
This also implies that the observed polarisation is likely associated with the irradiated accretion disc or scattering of its radiation in the wind. 

This interpretation is consistent with the wavelength dependence of intrinsic polarisation degree, with the peak in {\it B}, as obtained using star 2 as a proxy for ISM polarisation (see the blue dashed lines in Fig.~\ref{fig:spec_intr}a). 
In the {\it R} band, there is a stronger contribution of the unpolarized emission from the hot flow.
The larger increase in the {\it R}-filter polarisation can be caused by the greater reduction of the {\it R}-band hot flow emission, which is expected in this scenario.

If, on the other hand, we accept the five-star average as the ISM polarisation estimate, the intrinsic polarisation has a minimum in $B$ (see the red solid line in Fig.~\ref{fig:spec_intr}a).
This minimum can be understood as a contribution of unpolarized Balmer lines to this band. 
The wavelength dependence of the PA, which shows a deviation in $B$ by about 10\degr (see Fig.~\ref{fig:spec_intr}b), can indicate the difference of local normal to the disc plane between the places where {\it B} and {\it V}/{\it R} emission is produced.

We note that the emission from the hot flow is produced by relativistic electrons injected by magnetic reconnection, for example, with their subsequent fast cooling in the magnetic field. 
This means that the hot flow model and the fast-cooling base-of-the-jet model would produce similar spectra if the size of the emission region and the magnetic field were similar.  
The only difference is the direction of escaping particles down to the black hole within the flow or up to the jet.
The latter cannot be probed so far.
Therefore, it is currently impossible to differentiate between these models if the considered region of fast cooling in the jet is contained within the hot flow (i.e. when the jet is anchored in the accretion flow).

The requirement that the photons are able to escape avoiding self-absorption within the source and to produce the mid-IR emission down to 0.1~eV determines its size of about 400  Schwarzschild radii \citep[see Eq. 13 in][]{VPV13} for a 10M$_{\sun}$ black hole.  
We conclude that the hot flow is a viable possibility to reproduce the unpolarised emission in the mid-IR to optical, while the polarized emission likely comes from the irradiated disc or scattering in the outflow.

\subsection{Comparison to V404~Cyg}

The exceptional brightness of MAXI~J1820+070 is comparable to the recent black hole transient V404~Cyg, for which several high-precision (accuracy higher than 0.1\%) polarimetric observations were performed \citep{TIE16,KBP17}.
The intrinsic polarisation measured using Dipol-2 instrument was found at a level below $\sim$$1$\%, which is similar to the polarisation measured for MAXI~J1820+070.
The wavelength dependence of the intrinsic polarisation in V404~Cyg, with the possible peak in $V$, is similar to that obtained for MAXI~J1820+070 using the star 2 estimation for the ISM polarisation.

The greatest difference is, however, that V404 Cyg was observed during a likely super-Eddington outburst, while MAXI~J1820+070 was in the hard state, at a luminosity of about 10\% of the Eddington luminosity. 
In V404 Cyg, there was  evidence for a strong slow equatorial  outflow, which was a likely source of polarisation. 
Scattering in a mildly relativistic outflow was also suggested as an alternative model. 
Here, for MAXI~J1820+070 the outflow model is also possible because of the presence of the steady radio emission that is often observed in the hard states of accreting black holes.
We can note that in both cases, a low polarisation implies that the magnetic field in the emission region is rather disordered if the observed radiation is produced by synchrotron radiation in a jet or a hot flow and/or if the Faraday rotation plays a significant role in decreasing the polarisation.
The  wavelength dependence of the polarized emission implies a strong contribution of the wind or the irradiated accretion disc to the polarized flux.

\section{Conclusions}\label{sect:conclus}

We presented the results of the observational campaign studying {\it BVR} polarisation properties of the black hole X-ray binary candidate MAXI~J1820+070.
We found evidence for a low, but statistically significant polarisation, about 0.7\%--0.9\%, in the source direction, at PA$\approx$$50\degr$ in all three filters. 
This high accuracy is possible owing to the original design of the Dipol-2 instrument.

We found that the observed polarisation degree increases by about 0.1\% after MJD~58222 in the $V$ and $R$ filters with an additional change in PA by 4$\degr$ and 7.5$\degr$. 
In $B$, an enhancement of the polarisation at a $3\sigma$ level is seen. 
The moment when the polarisation started to significantly evolve coincides in time with the drop in the $V$ flux, suggesting that an increase in polarisation is associated with the decreasing flux of the non-polarized component.
We noted that at about the same date, the X-ray spectral index started to increase, suggesting a start of the transition to the soft state. 

To determine the interstellar component of polarisation, we performed a polarimetric study of the field stars and found the ISM polarisation degree of $\sim$0.6\%--0.8\% at PA$\approx$$65\degr-80\degr$, depending on the way we estimate it. 
The resulting intrinsic source polarisation degree is at a level of $\sim$0.3\% to 0.7\% at PA$\approx$$10\degr-30\degr$.

We suggest that the jet or the hot flow contribute to the optical flux, but either of these components have to be unpolarized, implying a rather tangled magnetic field or a large role of the Faraday rotation that destroys polarisation in the ionised magnetized medium close to the black hole.  
The likely source of the intrinsic polarized emission is the outer irradiated accretion disc or the disc radiation that is scattered by the optically thin wind.

\section*{Acknowledgements}

This research has been supported by the Ministry of Science and Higher Education of the Russian Federation grant 14.W03.31.0021. 
We acknowledge support from the Academy of Finland grants 309308 (AV) and 295114 (JJEK), ERC Advanced Grant HotMol ERC-2011-AdG-291659 (SVB, AVB).
The DIPOL-2 was built in cooperation by the University of Turku, Finland, and the Kiepenheuer Institut fuer Sonnenphysik, Germany, with support from the Leibniz Association grant SAW-2011-KIS-7. We are grateful to the Institute for Astronomy, University of Hawaii for the observing time allocated for us on the T60 telescope.
The research has made use of MAXI data provided by RIKEN, JAXA and the MAXI team, the NASA/IPAC Infrared Science Archive, which is operated by the Jet Propulsion Laboratory, California Institute of Technology, under contract with the National Aeronautics and Space Administration, and the data from the European Space Agency (ESA) mission {\it Gaia} (https://www.cosmos.esa.int/gaia), processed by the {\it Gaia} Data Processing and Analysis Consortium (DPAC, https://www.cosmos.esa.int/web/gaia/dpac/consortium). 
Funding for the DPAC has been provided by national institutions, in particular the institutions participating in the Gaia Multilateral Agreement.
We acknowledge with thanks the variable star observations from the AAVSO International Database contributed by observers worldwide and used in this research.

%%%%%%%%%%%%%%%%%%%%%%%%%%%%%%%%%%%%%%%%%%%%%%%%%%



\bibliographystyle{aa}
%\bibliography{allbib} 
%
%\end{document}

%%%%%%%%%%%%%%%%%%%%%%%%%%%%%%%%%%%%%%%%%%%%%%%%%%

 %%%%%%%%%%%%%%%%%%%%%%%%%%%%%%%%%%%%%%%%%%%%%%%%%%

% Don't change these lines
%\bsp   % typesetting comment
\label{lastpage}
\end{document}